# Solid-State Maser with Microwatt Output Power at Moderate Cryogenic Temperatures

Yefim Varshavsky[1], Oleg Zgadzai[1] and Aharon Blank[1*]

[1] Schulich Faculty of Chemistry, Technion – Israel Institute of Technology, Haifa, 3200003, Israel

*Corresponding author. Email: ab359@technion.ac.il

**Abstract:** Solid-state masers are uniquely positioned to serve as ultra-low phase noise microwave sources due to their exceptionally low noise temperatures. However, their practical application has been historically limited by low output power and the need for deep cryogenic cooling. In this work, we present a novel design for a continuous-wave diamond-based maser oscillator operating at ~14.5 GHz and moderate cryogenic temperatures (~180 K), achieving output power levels exceeding –30 dBm (1 µW). This performance represents a two-orders-of-magnitude improvement over previous diamond or ruby-based maser oscillators. Our system integrates a high-$Q$ (~2460) compact metallic microwave cavity with optically pumped [111]-oriented NV-rich diamond crystals. The cavity supports efficient light coupling and thermal dissipation, enabling sustained high-power optical excitation (>1 W) using cost-effective green LEDs. We demonstrate stable maser operation with good spectral quality and validate its output through both frequency- and time-domain analysis with phase noise data when operated in "free running" mode. Additionally, we provide phase noise estimations based on Leeson's model and show that, when coupled to a high-Q external resonator, such masers could approach thermally limited phase noise levels. These predictions suggest strong potential for diamond masers to outperform traditional ruby-based or similar maser systems, especially given their ability to operate at higher temperatures using rugged, He-free Stirling coolers. Despite current limitations related to frequency stability and jitter, this work establishes diamond-based masers as promising candidates for next-generation ultra-low phase noise microwave oscillators. Further engineering optimization—particularly in field stability, thermal regulation, and feedback locking—will be key to unlocking their full potential.

Solid-state masers (microwave amplifiers by stimulated emission of radiation), originally developed in the 1950s, are devices that can generate and amplify microwave radiation with very low additional thermal noise [1]. These systems traditionally rely on paramagnetic species, mainly ions embedded in a crystal, such as ruby ($Cr^{3+}$ in an $Al_2O_3$ crystal). Under an external static magnetic field, the paramagnetic species in the solid-state maser exhibit discrete energy levels. Various energy pumping schemes can be used to achieve a state of population inversion, where more electrons occupy a higher energy level than a lower one. Under such conditions, oscillations can occur, generating a microwave signal at the frequency corresponding to the energy level difference. Alternatively, with the right design, incoming microwave (MW) radiation can stimulate the emission of additional microwave radiation—thus amplifying the incoming signals. While the original solid-state masers could operate efficiently only at low cryogenic temperatures (~1 K) [2], recent developments in diamond-based masers have demonstrated operation at much higher temperatures [3,4], including room temperature [5-9]. These recent experiments have utilized masers both as microwave sources and as amplifiers. One major limitation of current designs is their *low power output* in continuous wave mode of operation. For example, when used as an oscillator, the best performance to date of diamond-based masers has shown an output power of only ~-54.1 dBm (~$3.9 \times 10^{-9}$ W) [8]. Ruby-based maser oscillators at ~ 1.5 K also have power levels on the range of ~$10^{-9}$ W at most [10]. Such power levels are too low for masers to serve as practical oscillators in applications requiring ultra-low phase noise, and they also restrict their use as amplifiers due to low saturation input power. The latter limitation is evident, while the former can be explained as follows: At room temperature, the thermal noise is approximately -177 dBm/Hz [11]. Therefore, to achieve a phase noise level of -160 dBc (160 dB below the carrier power), which is about the state-of-the-art at a 1 kHz frequency offset from the carrier for ~10 GHz frequencies [12], the corresponding microwave source must have an output power of at least -17 dBm to remain dominant above the thermal noise. At lower temperatures, the thermal noise limit is reduced, but a significant reduction would require deep cryogenic temperatures, which would greatly complicate the oscillator system. Even with such cooling, achieving sufficiently high power levels remains a challenge.

Here, we present a new design for a diamond-based solid-state maser operating at approximately 14.5 GHz, with a continuous-wave output power of up to -30 dBm (1 µW) at modest cryogenic temperatures of 180 K. This design incorporates a unique metallic microwave resonator with a relatively high-quality ($Q$) factor, allowing light access while ensuring efficient coupling. Additionally, it operates in conjunction with relatively large and unconventional diamond samples, designed for optimal thermal dissipation. The achieved microwave output power levels exceed those of previous designs by more than two orders of magnitude, providing a strong foundation for the development of future diamond-based ultra-low phase noise maser microwave oscillators.

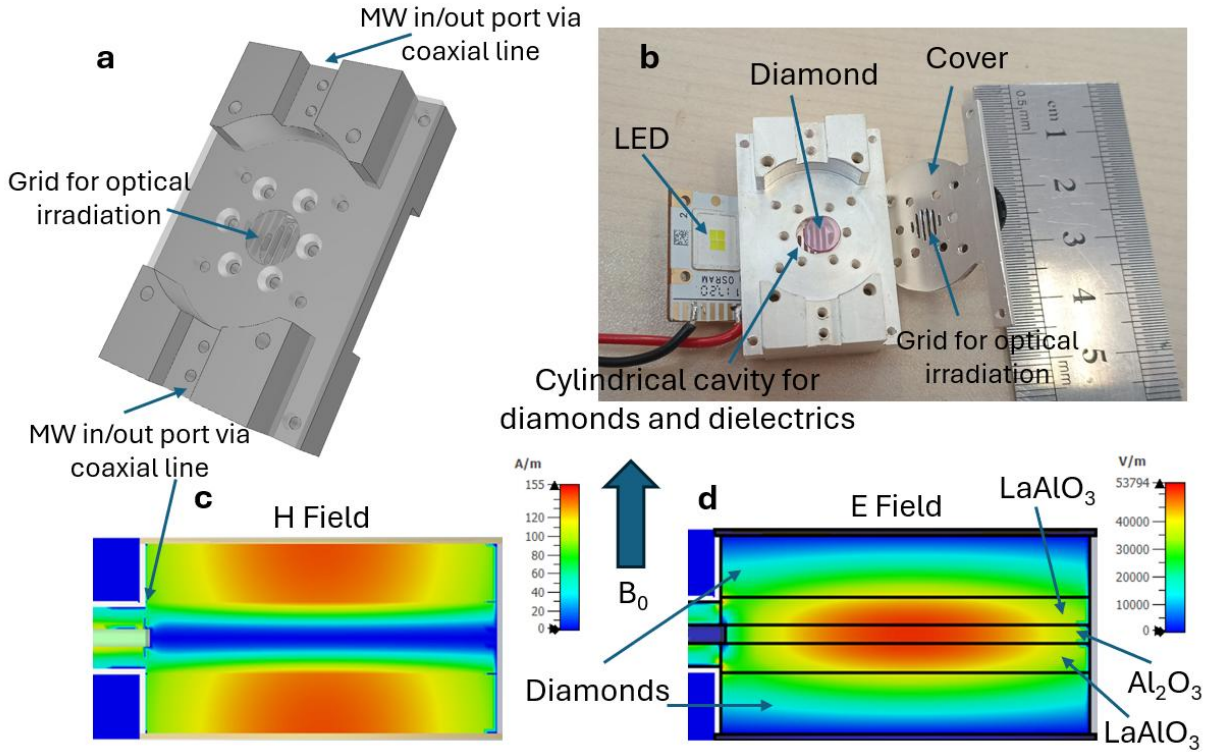

**Figure 1:** **The maser device.** (a) Schematic drawing of the custom-made maser resonator, featuring two ports for microwave (MW) input and output, as well as two optical windows for light irradiation from both sides. (b) Photograph of the resonator in its open state. It is made of high purity aluminium, coated with silver. The central cavity has a diameter of 6 mm and holds a stack of two round diamonds (each 1 mm thick), with two LaAlO$_3$ single crystals (0.5 mm thick) and one Al$_2$O$_3$ single crystal (0.3 mm thick) placed between them (not shown in the photo, see panel d for details). There are two optical windows (slits in the metallic design) that enable two green LEDs (only one is shown) to efficiently illuminate the two diamonds from both sides. (c) Simulation of the MW magnetic field distribution for the cavity's resonant mode, with the diamonds, LaAlO$_3$, and Al$_2$O$_3$ inside. The MW magnetic field is predominantly concentrated within the diamond crystals. (d) Same as (c), but for the MW electric field distribution, which is primarily concentrated within the LaAlO$_3$ and Al$_2$O$_3$ crystals. The static magnetic field ($B_0$) is perpendicular to the resonator surface and is along the metallic cylindrical cavity axis.

Figure 1 presents our maser device design (for a detailed drawing, see the Supplementary Material). The design is based on a unique microwave resonator that incorporates both metallic and dielectric components, facilitating high-efficiency optical pumping schemes. This design must comply with several key requirements:

1. *Overcoming the Maser Threshold:* The most critical requirement is that the device must exceed the maser threshold, meaning the net power gain due to population inversion must be greater than the resonator losses. Quantitatively, this condition can be expressed as:

$$Q_{th} = Q_m \equiv \frac{1}{\eta \chi''} = \frac{8}{\eta \Delta n \mu_0 \gamma^2 \hbar T_2^*}, \quad [1]$$

where $Q_{th}$ is the minimum value for the quality factor of the resonator required to sustain maser self-oscillating action, $Q_m$ is the quality factor due to the maser's gain medium (referred here in its absolute value, but actually having a negative value), and $\chi''$ is the microwave magnetic susceptibility of the NVs in the diamond at resonance. The terms $\mu_0$ and $\gamma$ are physical constants (free space permeability and electron gyromagnetic ratio factor, respectively), while $T_2^*$ is related to the inverse spectral linewidth of the NV$^-$, $\Delta f$, by the expression $T_2^* = 1/\pi \Delta f$, and $\Delta n$ is the population inversion difference between the upper and lower energy levels of the NV$^-$ transition used for maser amplification, per unit of volume. The term $\eta$ denotes the filling factor and is defined as $\eta = \int_{Diamond} |B_1^t|^2 \Big/ \int_{Resonator} |B_1|^2$, where the nominator involves the size of the MW magnetic field component, $B^t{}_1$, tangential to the direction of the static magnetic field, $B_0$, and the denominator includes the size of the total MW magnetic field $B_1$. One possible approach to exceeding the maser threshold is to use ultra-high-Q (>20,000) resonators [6,8,9,13]. However, maintaining such a high $Q$ often requires limiting the diamond size and NV density to minimize losses, which results in low $\eta$ and $\Delta n$. Additionally, such designs often suffer from poor thermal management, as high-$Q$ dielectric structures are typically kept away from metallic walls to prevent microwave losses. In our design, however, being closer to the metallic walls can help dissipate heat generated by optical pumping.

2. *Employing Diamonds with High NV Concentration:* To achieve high-power maser operation, the design must incorporate diamonds with a relatively high NV concentration, ensuring sufficient population inversion while emitting the required output power.
3. *Supporting High-Power Optical Excitation (~1 W):* Due to the inherently low energy efficiency of the maser, sustaining high-power optical excitation is crucial. Ideally, one green-light photon would produce one microwave photon, meaning that only about ~$2.5 \times 10^{-5}$ (equal the ratio of frequencies ~$14.5 \times 10^9 / 5.6 \times 10^{14}$) of the optical energy can be converted into a microwave signal. Consequently, with 1 W of optical excitation power, accounting for unavoidable losses and mismatches, the expected output power should be at least 1 µW, assuming a sufficient number of NV centers in the structure.

The resulting design shown in Fig. 1 complies with all these requirements, it features a compact metallic cavity with native relatively high unloaded $Q$ of >2000 into which we tightly place a stack of two large diamonds and dielectric materials that, due to the unique MW mode design, have minimal effects of the cavity $Q$. The same goes for the optical windows that are placed in a way that would have very minor effect on the structure $Q$ and still would enable every efficient light excitation of the diamonds by simple and affordable green LEDs (see Supplementary Material for details). Furthermore, the tight structure provides an efficient heat dissipation path. The design is also very slim, enabling to position it in a fairly compact permanent magnet (although this feature was not exploited in the present work). The dimensions of the metallic cavity were chosen to support the largest diamonds we could obtain, and the calculated filling factor was found to be $\eta = 0.6$. The unloaded quality factor of the resonator is ~2460 (see Supplementary Material for $S_{11}$ plot). The diamonds used in this work (see Supplementary Material for details) are [111] oriented, so that the static magnetic field, which is directed along the cavity cylindrical axis is also directed along this NV axis for optimal energy levels pumping efficiency [14]. Their NV concentration was measured to be ~20 ppm.

The maser oscillator was initially tested using an electromagnet as the source of the static magnetic field, which exhibited relatively large field fluctuations. The emitted microwave signal was monitored in the frequency domain using a spectrum analyzer, with all measurements conducted at a temperature of 180K. Figure 2a presents a typical MW signal captured in one of the spectrum analyzer scans. This scan demonstrates an unprecedentedly high output power of nearly 1 µW, along with fairly good spectral quality. However, in other scans, a more chaotic pattern can be observed, sometimes featuring multiple spectral lines, as shown in Figure 2b. This instability arises from electromagnet field fluctuations, which in our setup can easily vary by a few tenths of a Gauss, translating to more than 1 MHz of spectral shifts. Additionally, the native spectral linewidth of our sample is relatively broad (6.4 MHz, as shown in Figure S3), further complicating the spectral stability and contributing to even larger spectral fluctuations. To better understand this behaviour, we analysed the spectral variations in the time domain by feeding the maser MW signal into our pulsed ESR spectrometer, recording it as a standard ESR signal with downconversion to 50 MHz. Figure 2c shows a representative time-domain trace for periods with relatively stable spectral output, whereas Fig. 2d displays a typical signal during periods of faster frequency fluctuations.

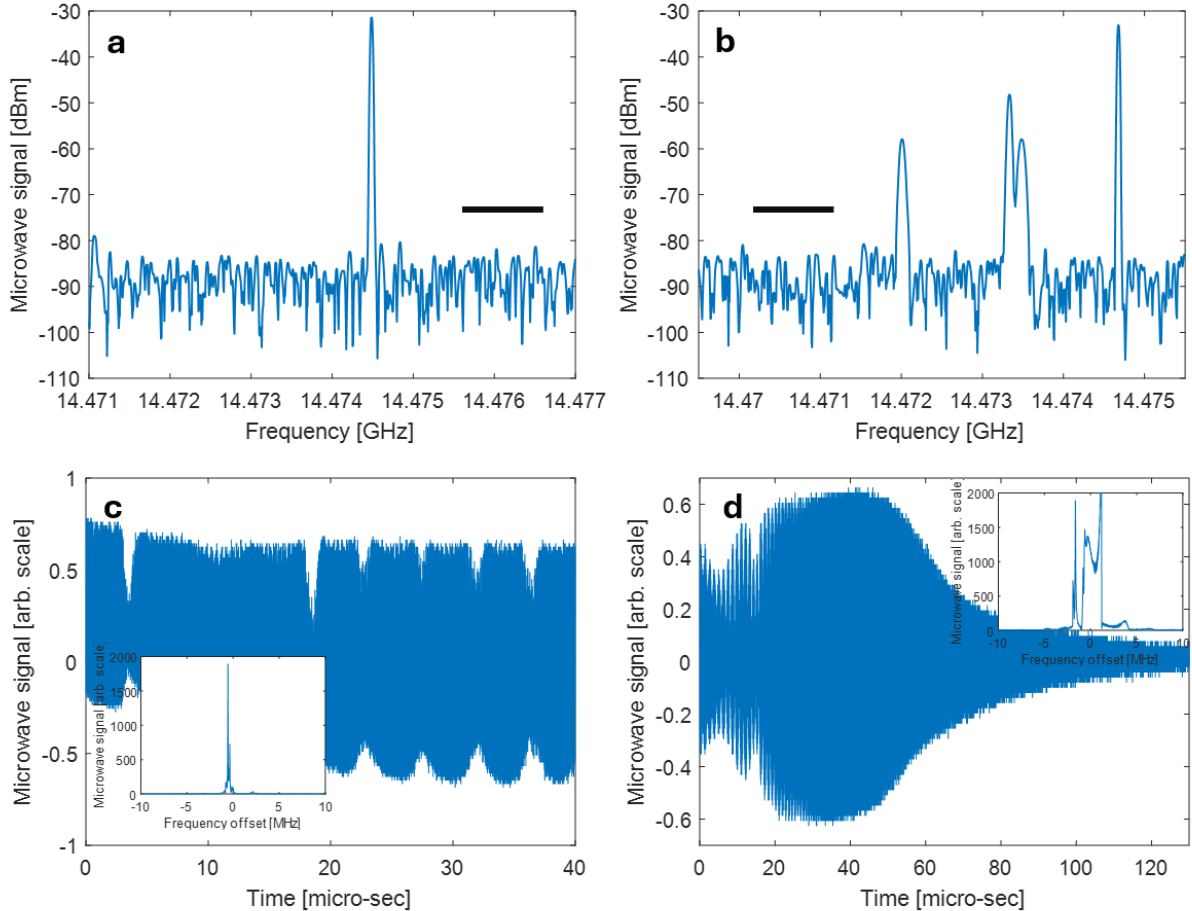

**Figure 2: Microwave signal characteristics of the maser inside an electromagnet.** (a) Spectrum analyser trace of the maser microwave (MW) signal acquired with a 6 MHz span, 1001 points, 2.33 ms sweep time, and 56 kHz resolution bandwidth, under 2.5 W optical excitation. The horizontal black scale bar represents 1 MHz. (b) Same as (a), but acquired during periods of reduced magnetic field stability. (c) Representative time-domain trace of the maser MW signal during a relatively stable period of the electromagnet. Inset: corresponding spectral domain view of the same signal. (d) Same as (c), but acquired during a period of reduced magnetic field stability.

Following this, we placed the maser resonator inside a superconducting magnet operated in persistent mode, which significantly reduced static magnetic field fluctuations compared to the previously used electromagnet. This resulted in improved spectral stability. Figure 3a presents a representative spectrum of the maser signal acquired under these conditions using a spectrum analyser (all measurements were conducted at 180 K). In contrast to measurements taken with the electromagnet, we generally observed good spectral purity in most scans, with output power slightly exceeding –30 dBm (~1 µW). However, the maser signal continued to exhibit frequency jitter, as evident from Figure 3b. This figure shows several scans acquired using the analyser's "max hold" mode, which records the maximum signal detected at each frequency across many scans. The corresponding time-domain signal (Figure 3c) further illustrates this frequency instability.

Following this, since our primary goal in increasing the maser's output power was to evaluate its potential as a low phase-noise microwave source, we attempted to directly measure the phase noise using the spectrum analyser in phase noise mode. Unfortunately, when the maser was operated at full optical excitation power (~1.2 W), the large frequency jitter—approaching 1 MHz (Figure 3b)—prevented the spectrum analyser from locking onto the maser frequency, thereby hindering accurate phase noise measurements. To mitigate this issue, we operated the maser close to its threshold, using a reduced optical excitation power of ~440 mW. Under these conditions, the frequency jitter decreased substantially to ~10 kHz, as oscillations could occur only near the peak of the NV ensemble's spectral line (Figure S3). However, this also reduced the maser output power to approximately –70 dBm, making direct phase noise measurements challenging due to thermal noise interference. We overcame this limitation by pre-amplifying the maser signal using a very low-noise amplifier (Model LNF-LNC6_20A, Low Noise Factory, Sweden) before going into the spectrum analyser. The resulting phase noise spectrum of the free-running maser, operated at minimal optical excitation power and with LNA amplification, is shown in Figure 3d.

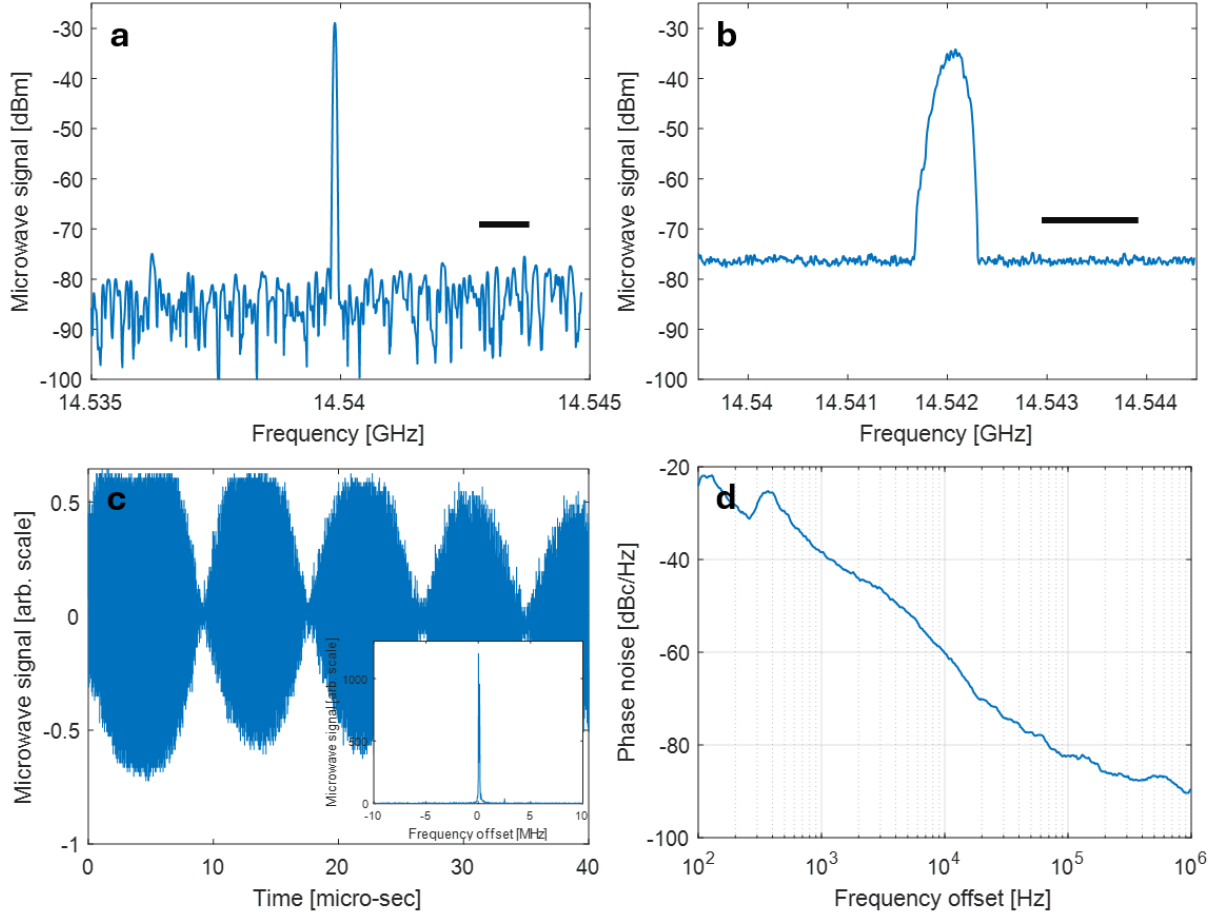

**Figure 3: Microwave signal characteristics of the maser inside a superconducting magnet.** (a) Spectrum analyser trace of the maser microwave (MW) signal acquired with a 10 MHz span, 1001 points, 1.47 ms sweep time, and 91 kHz resolution bandwidth, under 1.2 W optical excitation. The horizontal black scale bar corresponds to 1 MHz. (b) Spectrum analyser data acquired using the "max hold" mode, showing accumulated frequency jitter of the maser MW signal over multiple scans. Acquisition parameters: 5 MHz span, 1001 points, 2.13 ms sweep time, and 47 kHz resolution bandwidth, under 1.2 W optical excitation. The horizontal black scale bar corresponds to 1 MHz. (c) Representative time-domain trace of the maser signal during a period of relative field stability. Inset: corresponding spectral domain representation of the same signal. (d) Phase noise spectrum of the free-running maser, measured at low excitation power (~440 mW) and following signal amplification using a low-noise amplifier.

Our results show for the first time that a free-running solid-state maser, based on diamond NVs, can reach output power levels of more than 1 μW at very moderate cryogenic temperatures. Such power levels are well above what was achieved in previous designs for diamond masers and, in fact, to the best of our knowledge, higher than what was demonstrated for any other solid-state maser of similar size in the past at all temperatures. (For example, ruby maser has a typical power output of ~$10^{-7}$ W/cm³ [15]). Theoretically, at this power level of more than –30 dBm and a temperature of 180 K, the ultimate thermally limited phase noise level is ~–150 dBc. However, our measured phase noise was far from these values. We attribute this to the relatively broad linewidth of our NV transition and, mainly, to the lack of any high-Q resonator connected via a feedback loop for locking and stabilizing the maser frequency. Such a high-Q resonator is typical and vital for producing a low-phase-noise microwave source [16]. Quantitatively, one can analyze this situation using Leeson's equation, which provides a good estimation of the expected phase noise of a system with an amplifier (our maser, in this case) coupled to a high-Q factor external resonator, when the amplifier noise characteristics and other relevant parameters are known [17]:

$$L(f_m) = 10\log\left[\frac{k_B T_{sys}}{2P_{in}}\left(1+\left(\frac{f_0}{2Q_{ext}f_m}\right)^2\right)\left(1+\frac{f_c}{f_m}\right)\right], \qquad [2]$$

Where $P_{in}$ is the input power before the maser amplifier, $f_0$ is the operating frequency of the oscillator, and $Q_{ext}$ is the loaded quality factor of the oscillator (which can be approximated by the loaded $Q$ of the external resonator). $f_m$ denotes the offset frequency from the carrier, and $f_c$ is the flicker corner frequency (representing the 1/f noise contribution). Note that in the standard form of Leeson's equation, one typically finds the factor $Fk_BT$ rather than simply $k_BT_{sys}$, where $F$ is the noise figure of the amplifier and $T$ is the temperature. However, in the case of a maser, its equivalent noise temperature, $T_{sys}$, can be much lower than the ambient temperature. This is where the maser reveals its full potential as a low-phase-noise microwave source—assuming its saturation power is not too low. Figure 4 presents preliminary estimates of phase noise, based on Eq. [2], demonstrating the promising future of diamond-based masers as useful low-phase-noise microwave sources. These estimates consider operation at ~77 K and 50 K—temperatures that are still achievable using He-free, rugged Stirling cryogenic coolers. The "Nominal" values correspond to the performance achieved in this work, assuming coupling to a high-$Q$ resonator with $Q_{ext}=2\times10^6$, which is feasible at 77 K [18]. The system noise temperature $T_{sys}$ is estimated here at 4.5 K, using the standard expression for the amplifier noise temperature of a single-port maser amplifier $T_{sys} = |T_m| \frac{1}{1-\varepsilon} + T_0 \varepsilon$ [3]. This estimate assumes an ambient temperature $T_0$=77 K, a spin temperature approaching the quantum limit $T_m \approx 0.65$ K [3], and coupling factor $\varepsilon$=0.05, implying that the maser resonator quality factor ($Q$~2000 in our case) is approximately 20 times larger than $Q_m$ Eq. [1]. The "Ultimate" values represent potential future performance, assuming a tenfold increase in saturation power and, crucially, coupling the amplifier to an ultra-high-$Q$ sapphire resonator with $Q_{ext}$~$10^8$, operating at 50 K [19]. The expected performance of diamond-based masers is compared with that of ruby-based masers operating at ~2.6 GHz, in combination with ultra-high-$Q$ sapphire resonators housed in superconducting shields at ~1.5 K [20]. These results are similar to those obtained more recently for sapphire-based masers operating at ~8 K, producing X-band microwave sources with extremely low phase noise [21].

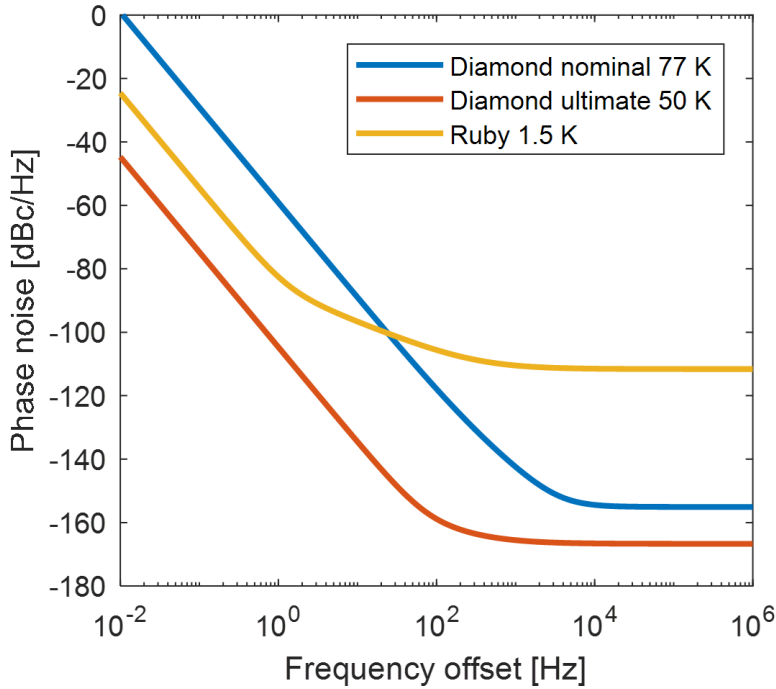

**Figure 4: Expected phase noise of diamond-based measure at ~77 K, compared to ruby-based maser at ~1.5 K.** (a) Nominal calculated situation with $T_{sys}$=4.5 K, $P_{in}$=0.1 µW, $Q$=2×$10^6$, $f_0$=14.5 GHz, $f_c$=300 Hz. (b) Ultimate calculated results for $T_{sys}$=3.1 K, $P_{in}$=1 µW, $Q$=1×$10^8$, $f_0$=14.5 GHz, $f_c$=300 Hz. (c) Calculated results for ruby maser, based on parameters of the system described in [20] that show similarity to the experimental data. $T_{sys}$=10 K, $P_{in}$=$10^{-11}$ W, $Q$=1×$10^9$, $f_0$=2.6 GHz, $f_c$=300 Hz.

It is concluded that diamond-based masers can produce relatively high output power, enabling them to serve as viable and relatively simple sources of ultra-low phase noise microwave signals. Their key advantage over ruby-based masers lies in their ability to operate at significantly higher ambient temperatures, eliminating the need for liquid helium cooling, while still maintaining comparable—or even superior—overall equivalent system noise temperatures. Furthermore, diamond masers can sustain higher output power within compact form factors, which is critical for achieving low phase noise performance at relatively large frequency offsets from the carrier. When compared to other microwave oscillator technologies, it becomes evident that maser-based systems in general—and diamond-based systems in particular—may be especially attractive in the frequency offset regime below approximately $10^4$ Hz [22]. However, substantial engineering work is still required to fully realize the potential of these systems. In addition to coupling to high $Q$ external resonator, this would include further optimization of diamond doping and maser resonator geometry, suppression of low-frequency flicker noise, stabilization of temperature under optical illumination, and the maintenance of a stable static magnetic field.

## Acknowledgments:

We thank IAI ELTA Systems for its support in this research. This project was funded by the Israel Science Foundation (ISF), Grant No. 1357/21, and the Israel Innovation Authority, Grant No. 77540.

Supplementary Material for

# Solid-state maser with micro-watt output power at moderate cryogenic temperatures

Yefim Varshavsky[1], Oleg Zgadzai[1] and Aharon Blank[1*]

[1] Schulich Faculty of Chemistry, Technion – Israel Institute of Technology, Haifa, 3200003, Israel

*Corresponding author. Email: ab359@technion.ac.il

## S1. The microwave resonator mechanical structure

The mechanical structure of the microwave resonator is provided in Fig S1. It is made of high purity aluminum (99.999%) from Advent Research Material LTD, to minimize magnetic impurities and then coated with silver (~ 8 μm with nickel supporting layer) after production.

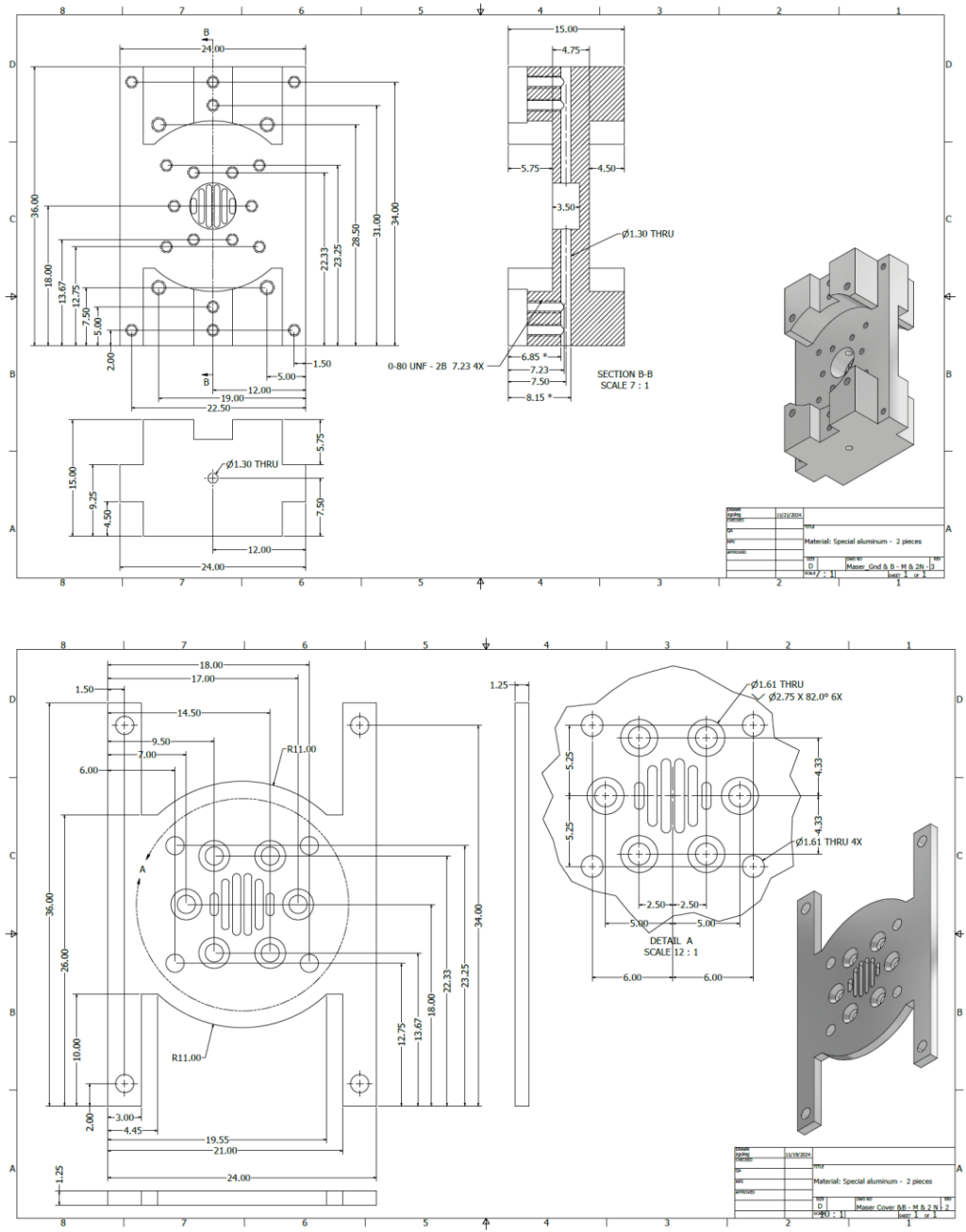

**Figure S1: The maser resonator mechanical structure.** (top) The main structure with the cavity to hold the diamonds. (bottom) Cover to close the structure and enable light irradiation through slits.

## S2. Green light source for the NVs electronic levels' pumping

We utilized relatively simple, high-power flat green LEDs, as shown in Figure 1 in the main text. These LEDs are model LE CG P2AQ from Osram Ltd., with a predominantly green irradiation spectrum peaking at approximately 512 nm. They are powered by an Agilent E3644A power supply, which has a current noise ripple of ~1% at a nominal current of 0.06–0.5 A per LED and an operating voltage of 10.5–12 V. The measured optical power output for each LED was ~220 mW at 60 mA, ~400 mW at 120 mA, and ~1.2 W at 400 mA. To avoid disturbances in the static magnetic field at the diamonds' position, the LEDs were positioned outside the resonator at a distance of ~4 mm from the diamond samples.

## S3. Microwave properties of the resonator

The resonator has two microwave (MW) ports, as shown in Figures 1 and S1. These ports allow for the insertion of a 0.047" semi-rigid coaxial copper cable, with a stripped shield of ~3 mm at its distal end. The coupling between the resonator and the external transmission line can be controlled by adjusting the insertion depth of the cable, allowing for a range from highly undercoupled to critical coupling and overcoupling. Figure S2 presents the measured reflection coefficient ($S_{11}$) of the resonator when it is close to critical coupling and only one port is connected. The loaded Q-factor is measured to be ~1230, indicating that the unloaded Q-factor of the resonator is ~2460.

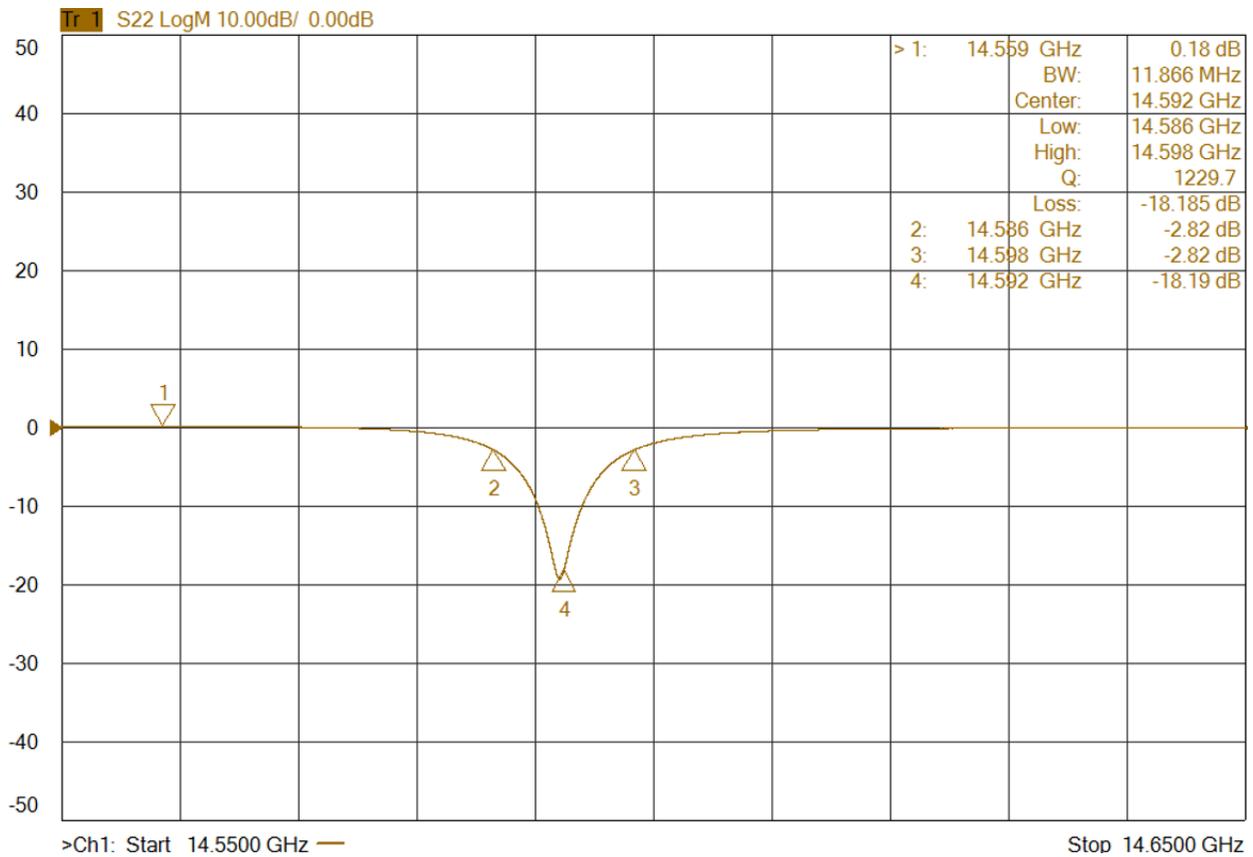

**Figure S2:** The maser resonator reflection coefficient, as measured by vector network analyzer.

## S4. The diamond samples

Diamond samples were purchased from Super More Hard Company (China) and synthesized using the high-pressure high-temperature (HPHT) technique. The diamonds had a substitutional nitrogen concentration of approximately 40 ppm. To generate nitrogen-vacancy (NV) centers, the samples underwent electron-beam irradiation at Golan Plastic Company (Israel) with a dose of ~$2\times10^{19}$ electrons/cm² at an electron energy of 2.65 MeV. Following irradiation, the diamonds were annealed in an argon atmosphere—initially for 4 hours at 800°C, followed by an additional 2 hours at 950°C. This process yielded an NV center concentration of approximately 20 ppm, as estimated by comparing the NV electron spin resonance (ESR) signal to that of calibrated reference samples. The ESR spectrum corresponding to the $|0\rangle \leftrightarrow |-1\rangle$ transition of the [111]-oriented NV centers is shown in Fig. S3.

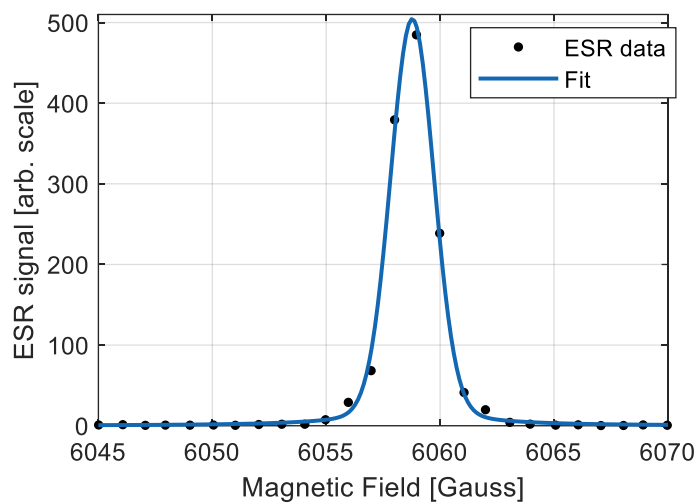

**Figure S3: Field-swept echo ESR spectrum for the NV.** Signal was measured using Hahn echo sequence with long 1000ns π/2 and π pulses and interpulse separation (τ) of 500 ns. Fit is made to mixed Gaussian and Lorentzian line shape with the resulting FWHM of 2.3G (~6.4 MHz).